# Inhomogeneity in barrier height at graphene/Si (GaAs) Schottky junctions

D. Tomer[1*], S. Rajput[1], L. J. Hudy[1], C. H. Li[2], and L. Li[1]

[1]Department of Physics, University of Wisconsin, Milwaukee, WI 53211, USA

[2]Naval Research Laboratory, Washington, DC 20375, USA

**Abstract**

Graphene interfaced with a semiconductor forms a Schottky junction with rectifying properties, however, fluctuations in the Schottky barrier height are often observed. In this work, Schottky junctions are fabricated by transferring chemical vapor deposited monolayer graphene onto n-type Si and GaAs substrates. Temperature dependence of the barrier height and ideality factor are obtained by current-voltage measurements between 215 and 350 K. An increase in the zero bias barrier height and decrease in the ideality factor are observed with increasing temperature for both junctions. Such behavior is attributed to barrier inhomogeneities that arise from interfacial states as revealed by scanning tunneling microscopy/spectroscopy. Assuming a Gaussian distribution of the barrier heights, mean values of 1.14±0.14 eV and 0.76±0.10 eV are found for graphene Si and GaAs junctions, respectively. These findings provide insights into the barrier height inhomogeneities in graphene based Schottky junctions, essential for the integration of graphene with Si and GaAs device architectures.

[*]Correspondence and requests for materials should be addressed to D. T. (dtomer@uwm.edu)

Graphene (Gr), a semimetal with linear energy dispersion[1], forms Schottky junction when interfaced with a semiconductor[2]. Unlike conventional metal-semiconductor junction, the graphene variant offers the unique advantage of a tunable Fermi level[3-5], and thus Schottky barrier height (SBH)[5,6], which enables novel device applications such as three terminal transistors with up to $10^6$ on/off ratio[5,7] and tunable reverse biased sensors[8]. The barrier height that is key to the performance of these devices is typically calculated by thermionic emission theory, which inherently assumes a perfect homogeneous junction interface[2,9-11]. Large variations in SBH, however, have been reported[2,6,9,12]. For example, SBH between 0.25 and 0.92 eV has been observed for Gr/Si Schottky junction[2,6,9,12]. These variations could be caused by either extrinsic effects such as the number of graphene layers in the junction[9,13] or intrinsic spatial inhomogeneity due to substrate steps[14] and graphene ripples and ridges[15,16], or the formation of electron and hole puddles in graphene[17,18]. The impact of graphene ripples on the SBHs of Gr/SiC Schottky junctions has been confirmed recently by scanning tunneling microscopy/spectroscopy (STM/STS)[15] and temperature dependent I-V measurements[16].

In this work, the effect of electronic inhomogeneities at monolayer Gr/Si and Gr/GaAs Schottky junctions on SBHs are investigated using STM/STS and temperature dependent I-V measurements between 215-350 K. Non-ideal behavior such as increase in zero bias SBH and decrease in ideality factor with increasing temperature are observed, which is attributed to the existence of a Gaussian distribution of the SBHs. This analysis yields mean SBHs of 1.14 and 0.76 eV for Gr/Si and Gr/GaAs junctions with standard deviations of 0.14 and 0.10 eV, respectively.

The Schottky junctions were fabricated by transferring chemical vapor deposited

(CVD) graphene onto n-type Si(111) and GaAs(100) with carrier densities $N_d \sim 10^{17}$ cm$^{-3}$ and $\sim 10^{16}$ cm$^{-3}$, respectively. The substrates were first cleaned with organic solvents followed by RCA procedure. To terminate the surfaces with hydrogen[19] and sulfur[20], the as-cleaned Si and GaAs wafers were dipped in 49% HF+40% NH$_4$F (1:7) and NH$_4$S (40%) solutions, respectively. The top electrode was formed by depositing Cr/Au (10/150 nm) on top of 100 nm SiO$_2$ grown on the H- and S-terminated Si and GaAs by electron beam evaporation (base pressure $\sim 2 \times 10^{-6}$ Torr). A square of 0.9 x 0.9 mm$^2$ is made for graphene contact. To form Ohmic back contact on GaAs, multilayer Au/Ni/AuGe (100/20/50 nm) was deposited by electron beam evaporation at room temperature (base pressure $\sim 2 \times 10^{-6}$ Torr), followed by annealing at T$\sim$ 400 $^0$C in forming gas. Conducting silver paste was used to make Ohmic back contact on Si. Finally, monolayer CVD graphene (Graphene Platform, Inc.) was transferred using the PMMA based method[21] onto the Si and GaAs substrates with prepatterned electrodes, with a quick 10 second dip into the NH$_4$F and NH$_4$S solutions right before the graphene transfer. Temperature dependent I-V measurements were carried out using a Keithley 2400 source meter between 215 and 350 K. STM/STS was carried out at liquid nitrogen temperature. The dI/dV tunneling spectra are acquired using lock-in detection by turning off the feedback loop and applying an AC modulation of 9 mV (r.m.s.) at 860 Hz to the bias voltage.

The surface morphology of transferred graphene is first characterized by STM. Figure 1(a) shows an image of graphene / Si substrate after annealing at $\sim$200 $^0$C in UHV for 30 min. Clearly evident is a non-uniform surface with vertical undulations of $\sim$0.5 nm over length scales of tens of nanometers (marked by a circle), likely due to roughness of the underlying Si substrate. Figure 1(b) is a close-up view showing the characteristic graphene honeycomb lattice that is continuous over these fluctuations. These features are similar to earlier STM studies of graphene ripples[15,16, 22,23], which are attributed to either graphene in contact with the underlying substrate (dark regions), or buckled up from it (bright regions).

The electronic properties of the graphene are further investigated by tunneling spectroscopy. Figure 1(c) shows spatially resolved dI/dV spectra taken across a ripple at locations marked in Fig. 1(b) for Gr/Si. All spectra exhibit two characteristic minima, one at zero bias ($E_F$) caused by phonon-assisted inelastic tunneling[24], and the other at negative bias marked by downward arrows attributed to the Dirac point ($E_D$), indicating n-type doping[25]. Moving from bright to dark to bright regions [Fig. 1(c)], while $E_D$ varies between 105 and 130 meV, no direct correlation is found to the topographic fluctuations, in contrast to the case of graphene transferred on SiC substrates[15,16]. In addition, atop the brightest regions (spectra 1-3 and 7,8) peaks also appear, as marked by upward arrows, possibly due to states arising from disorder from the partial hydrogen termination of the Si substrate).[26]

Similar features are observed for graphene / GaAs as shown in Fig. 2(a). Large scale corrugations of ~ 1 nm in height and hundreds of nm in width likely originated from substrate roughness. At the atomic scale, ripples ~ 0.35 nm in height are also seen (Fig. 2(b)). A series of dI/dV spectra, taken at positions 1-11 in Fig. 2(b), are shown in Fig. 2(c). While all spectra exhibit the similar phonon-assisted inelastic tunneling[24] at $E_F$, the Dirac point (marked by downward arrows) is now above $E_F$, indicative of p-type doping. Again, fluctuations in Dirac energy between 110 and 160 meV are also observed, but with no direct correlation is found with the undulation of the ripples. Likely substrate disorder induced states peaked at ~0.24 eV are again observed at some locations (spectra 1-3, 5).

The local fluctuations in the Dirac point lead to variation in carrier concentration ($\Delta n(p)$) that can be calculated by $\Delta n(p) = \frac{4\pi(\Delta E_D)^2}{(hv_f)^2}$, where $v_f$ is the Fermi velocity of graphene (~ c/300, where c is the speed of light) and $h$ the Plank's constant. This yields variations of $3.79 \times 10^{10}$ cm$^{-2}$ and $1.57 \times 10^{11}$ cm$^{-2}$ in electron and hole concentrations for Gr/Si and Gr/GaAs junctions, respectively.

These observations clearly indicate that graphene is prone to ripple formation when interfaced with Si and GaAs substrates, similar to CVD graphene transferred on hydrogen-terminated SiC substrates[15,16] and exfoliated graphene on SiO$_2$[23]. Interestingly, unlike the Gr/SiC junctions, the spatial variations in E$_D$ for both junctions do not follow the topographic fluctuations[15,16]. The local carrier fluctuations due to Dirac point variation nevertheless results in electron and hole puddles, similar to that of graphene / SiO$_2$[23]. As discussed below, this inherent spatial inhomogeneity in graphene can lead to fluctuations in the SBH determined by the temperature dependent I-V measurements.

Between 215 and 350 K, both Gr/Si and Gr/GaAs junctions show rectifying I-V, as shown in Fig. 3(a) and (b), respectively, suggesting the formation of Schottky diodes. The thermally excited transport across the junction follows the standard thermionic emission (TE) model[27]

$$I(T,V) = I_S(T)\left[\exp\left(\frac{qV}{\eta kT}\right) - 1\right] \qquad 1$$

where $V$ is the applied voltage, $q$ the electron charge, $k$ the Boltzmann's constant, and $\eta$ the ideality factor. The saturation current, $I_s$ (T), can be expressed as[27]

$$I_S(T) = AA^*T^2 \exp\left[\frac{-q\phi_{b0}}{kT}\right] \qquad 2$$

where $A$ is the diode area, $A^*$ the effective Richardson constant of the semiconductor, and $\phi_{b0}$ the zero bias SBH, which can be obtained from the extrapolation of $I_s(T)$ in the semi-log forward bias $ln(I)$-$V$

$$\phi_{b0}(T) = \frac{kT}{q}\ln\left(\frac{AA^*T^2}{I_S}\right) \qquad 3$$

The ideality factor, $\eta$, is a dimensionless parameter that accounts for any deviation from the standard TE theory ($\eta=1$ for an ideal junction), and can be calculated from the slope of the linear region of the forward $ln(I)$-$V$

$$\eta = \frac{q}{kT}\left(\frac{dV}{d(\ln I)}\right) \qquad 4$$

Temperature dependent forward bias $ln(I)$-$V$ plots of Gr/Si and Gr/GaAs junctions are shown in Figs. 3(c) & (d), respectively. At low bias voltages, both are linear over ~3-4 orders of current, and the deviation from the linearity is likely due to large series resistance[28] in both types of junctions. Figure 4(a) shows the temperature dependence of the zero bias SBH $\phi_{b0}$ and ideality factor $\eta$, calculated using equations (3) & (4), with a total diode area of 1.62 mm$^2$ (=2x 0.9 x0.9 mm$^2$), and Richardson constant $A^*$ of 1.12x10$^6$ Am$^{-2}$ K$^{-2}$ and 0.41x10$^4$ Am$^{-2}$K$^{-2}$ for n-Si[9,29] and n-GaAs[30], respectively. For Gr/Si, $\phi_{b0}$ increases from 0.66 to 0.82 eV and $\eta$ decreases from 2.62 to 1.66 from 215 to 350 K. A similar trend is also seen for Gr/GaAs, where $\phi_{b0}$ changes from 0.48 to 0.62 eV and $\eta$ varies from 1.88 to 1.44. This temperature dependence clearly deviates from the ideal TE theory, suggesting barrier inhomogeneities[31-33].

This is further supported by the analysis of the Richardson plot, $\ln(I_S/T^2)$ versus 1000/T (Fig. 4(b)). The deviation from the linearity below 275K indicates temperature dependent barrier height for both junctions[34]. Linear fitting of the data above ~275 K (dashed line) yields SBH of 0.47 and 0.36 eV, and $A^*$ of $1.04 \times 10^1$ and $8.72 \times 10^{-1}$ Am$^{-2}$K$^{-2}$ for Si and GaAs, respectively. The large deviation of $A^*$ from the known experimental values of $1.12 \times 10^6$ Am$^{-2}$K$^{-2}$ for Si[9,29] and $0.41 \times 10^4$ Am$^{-2}$K$^{-2}$ for GaAs[30] clearly indicates inhomogeneous SBHs due to potential fluctuations at the interface[35].

Assuming a Gaussian distribution of the barrier height, the deviation from ideal TE theory can be explained by the Werner model[36] that correlates the mean ($\phi_{bm}$) and apparent ($\phi_{ap}$) barrier height as follows

$$\phi_{ap}(T) = \phi_{bm} - \frac{q\sigma_s^2}{2kT} \qquad 5$$

where $\phi_{ap}(T)$ is the same as the experimentally measured $\phi_{b0}(T)$ and $\sigma_s$ is the standard deviation. The $\phi_{ap}(T)$ is plotted as a function of $q/2kT$ in Fig. 5(a), yielding a mean barrier height of 1.104 eV and $\sigma_s = 141$ mV for Gr/Si, and 0.76 eV and $\sigma_s = 98$ mV for Gr/GaAs.

The modified Richardson plot, [ $ln(I_s/T^2)-q^2\sigma_s^2/2k^2T^2$ ] vs $1000/T$, is shown in Fig. 5(b). Since

$$\ln\left[\frac{I_s}{T^2}\right] - \left[\frac{q^2\sigma_s^2}{2k^2T^2}\right] = \ln(AA^*) - \frac{q\phi_{bm}}{kT} \qquad 6$$

this plot should be a straight line with the the mean barrier height ($\phi_{bm}$) determined by slope and y-intercept [$ln(AA^*)$] that would directly yield $A^*$. The mean barrier height is found to be 1.15 eV for Gr/Si and 0.74 eV for Gr/GaAs, with corresponding Richardson constants of $1.14 \times 10^6$ and $0.27 \times 10^4$ A m$^{-2}$K$^{-2}$, respectively, in much better agreement with previously defines values.

Alternately, the barrier inhomogeneities can be considered by calculating the flat band barrier height $\phi_{bf}$, an intrinsic parameter given by[37]

$$\phi_{bf} = \eta\phi_{b0} - (\eta-1)\zeta \qquad\qquad 7$$

where $\zeta = (kT/q)ln(N_c/N_d)$, and $N_c = 2(2\pi m^* kT/h^2)^{3/2}$ is the effective density of states, and $N_d$ the donor concentration. Figures 6(a) & (b) show $\phi_{bf}$ as a function of temperature, where the dashed line is a linear fit with

$$\phi_{bf}(T) = \phi_{bf}(0) + \alpha T \qquad\qquad 8$$

where $\phi_{bf}(0)$ is the zero-temperature flat band barrier height and $\alpha$ the temperature coefficient. This yields a $\phi_{bf}(0)$ of 1.54 and 0.86 eV with $\alpha = 6.48\times10^{-4}$ and $1.09\times10^{-4}$ eV K$^{-1}$ for Gr/Si and Gr/GaAs, respectively. Clearly, the flat band barrier heights ($\phi_{bf}$) are not only always greater than the zero bias values ($\phi_{b0}$), but are also with a weak temperature dependence.

In addition, Eq. (7) also correlates the measured zero bias SBH $\phi_{b0}$ and ideality factor $\eta$. For homogeneous junction, $\eta=1$, thus $\phi_{bf} = \phi_{b0}$. For inhomogeneous junctions, $\eta$ is always greater than one. In the current case, since the magnitude of $\phi_{b0}$ is more than 10x greater than $\zeta$ for Gr/Si and ~5x greater for Gr/GaAs junctions, the term $\eta\phi_{b0}$ is much larger than $\zeta(\eta-1)$, hence $\phi_{b0} \sim \phi_{bf}/\eta$. Two conclusions can be drawn. First, the flat band barrier height $\phi_{bf}$ is always greater than the zero bias value $\phi_{b0}$ since $\eta$ is greater than one, consistent with experimental data. Second, in the limits when $\eta$ is small (or large), e.g., < 2.6, a linear relationship (with a negative slope) between $\phi_{b0}$ and $\eta$ can be approximated.

Indeed, as shown in Fig. 6(c), a linear relationship provides an excellent fit for the plot of the zero bias barrier heights as a function of ideality factor, similar to earlier studies of Gr/SiC[16] and Ag/Si[38] Schottky junctions. Extrapolation of the barrier height at the unity ideality factor leads to barrier heights of 0.98 eV and 0.72 eV for Gr/Si and Gr/GaAs junctions, respectively. These values are in good agreement with the mean SBH values obtained from the temperature dependent apparent barrier heights in Fig. 5(a) and modified Richardson plot in Fig. 5(b), confirming that transport across the Gr/Si and Gr/GaAs Schottky junctions is consistent with modified thermionic emission with a Gaussian distribution of barrier heights.

In summary, Gr/Si and Gr/GaAs Schottky junctions are investigated using atomic resolution STM imaging, *dI/dV* tunneling spectroscopy, and temperature dependent *I-V* measurements. The temperature dependent zero bias barrier height and ideality factor show clear deviations from the standard thermionic emission theory, which is explained by barrier height fluctuations caused by Dirac point inhomogeneities likely induced by semiconductor substrate disorder. Together with our earlier work on the Gr/SiC Schottky junctions, where Dirac point fluctuations correlate directly with topographical undulations of graphene ripples[15,16], these findings reveal two types of intrinsic inhomogeneities that can cause barrier height fluctuations in graphene / semiconductor Schottky junctions. Which mechanism dominates will depend on the nature of the semiconductor (e.g., polar vs non-polar), and/or the degree of disorder and roughness of the semiconductor surface.

**Acknowledgment**

Research supported by the U.S. Department of Energy, Office of Basic Energy Sciences, Division of Materials Sciences and Engineering under Award DE-FG02-07ER46228.

**Figure captions**

**Figure 1** (color online) (a) STM image of CVD graphene transferred onto n-Si substrate ($I_t$ = 0.1 nA, Vs = -0.65 V). (b) Atomic resolution STM image of graphene ripples ($I_t$ = 0.1 nA, $V_s$ = -0.3 V). (c) Spatially resolved dI/dV spectra taken at the locations marked in (b).

**Figure 2** (color online) (a) STM image of CVD graphene transferred onto n-GaAs substrate ($I_t$ = 0.1 nA, Vs = -0.3 V). (b) Atomic resolution STM image of graphene ripples ($I_t$ = 0.2 nA, $V_s$ = -0.3 V). (c) Spatially resolved dI/dV spectra taken at the locations marked in (b).

**Figure 3** (color online) (a) Temperature dependent *I-V* curves of Gr/Si Schottky junction between 215 and 350 K (inset: schematic diagram of the device). (b) Temperature dependent *I-V* curves of Gr/GaAs Schottky junction (inset: close-up view of forward bias current). (c) & (d) Temperature dependent semi-logarithmic forward bias *I-V* curves of Gr/Si & Gr/GaAs Schottky junctions, respectively.

**Figure 4** (color online) (a) Zero bias barrier height ($\phi_{b0}$) and ideality factor ($\eta$) as a function of temperature for Gr/Si and Gr/GaAs Schottky junctions. (b) Richardson plot, $ln(I_s/T^2)$ vs *1000/T* for Gr/Si and Gr/GaAs.

**Figure 5** (color online): (a) Apparent zero bias barrier height $\phi_{ap}$ vs as a function of $q/2kT$ for Gr/Si and Gr/GaAs junctions. (b) Modified Richardson plot, $ln(I_s/T^2)-q^2\sigma_s^2/2k^2T^2$ vs *1000/T* for the same junctions.

**Figure 6** (color online): Flat band barrier height $\phi_{bf}$ as a function of temperature for (a) Gr/Si and (b) Gr/GaAs junctions. (c) Zero-bias barrier height ($\phi_{b0}$) as a function of ideality factor ($\eta$) for the same junctions.

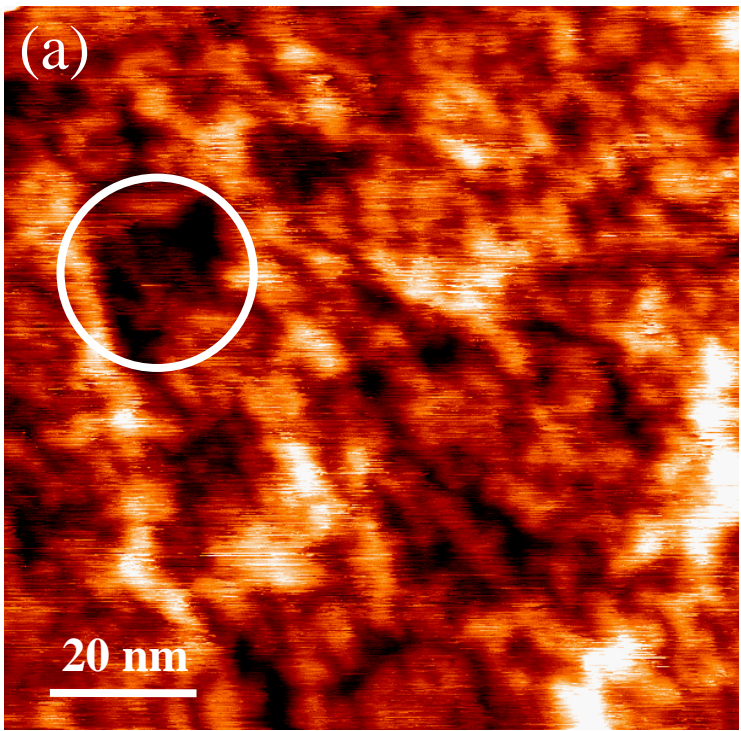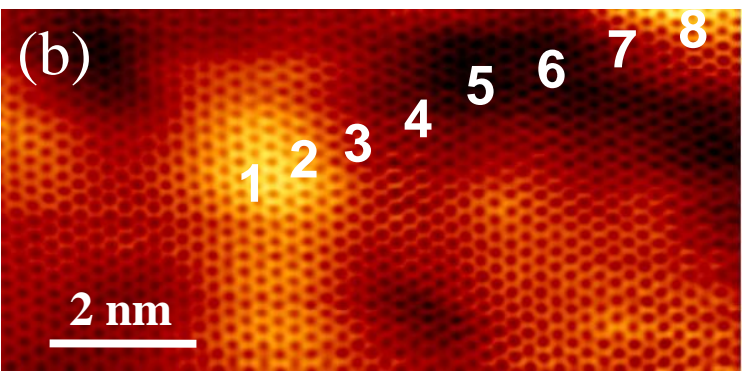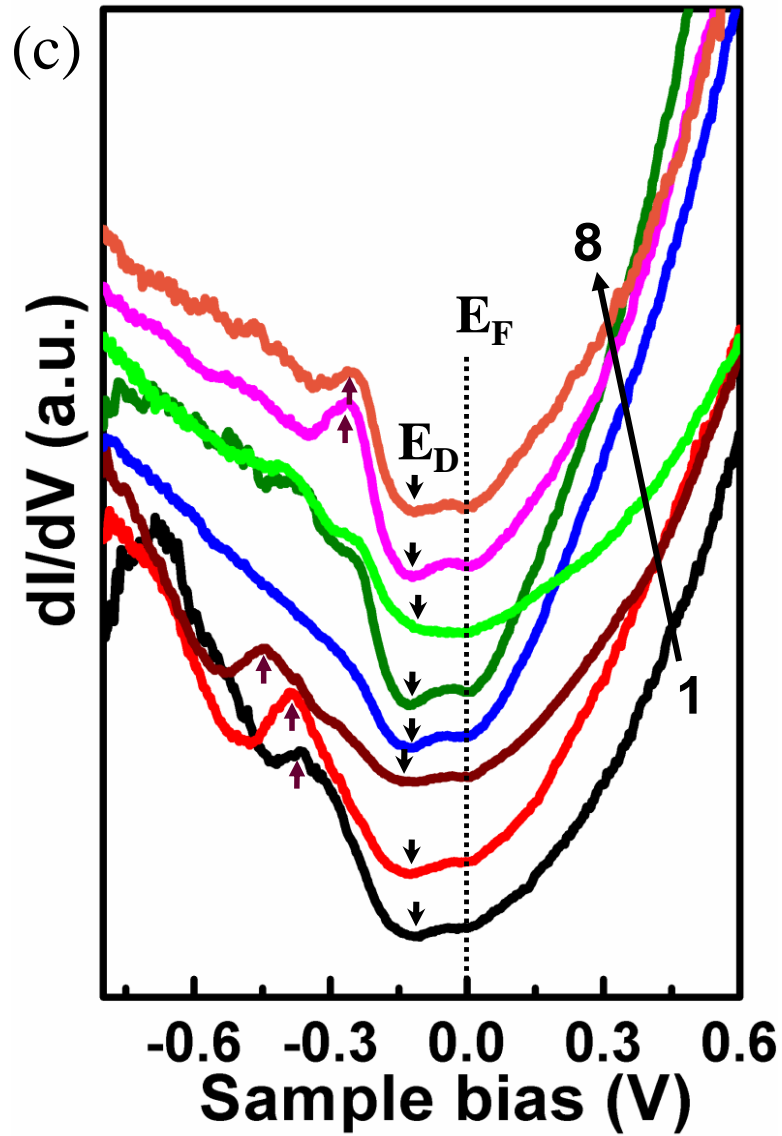

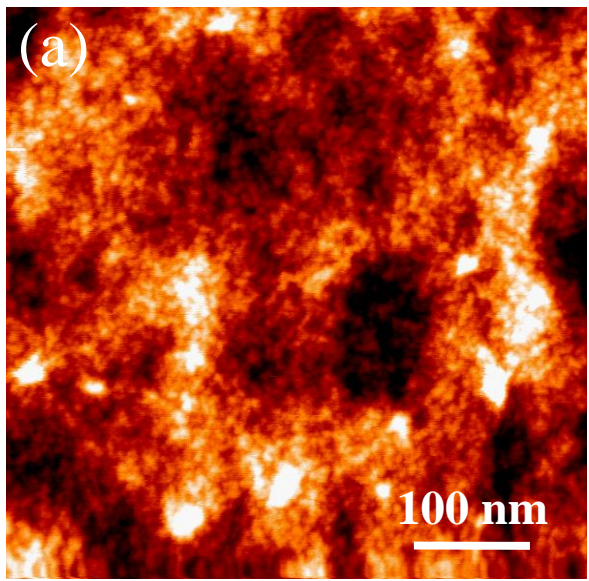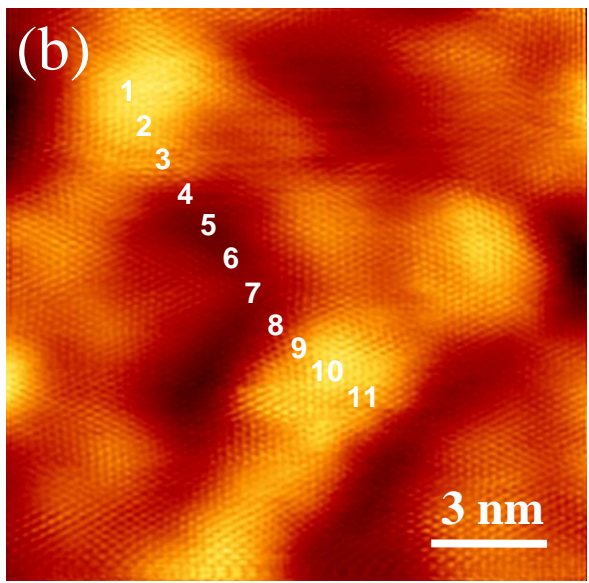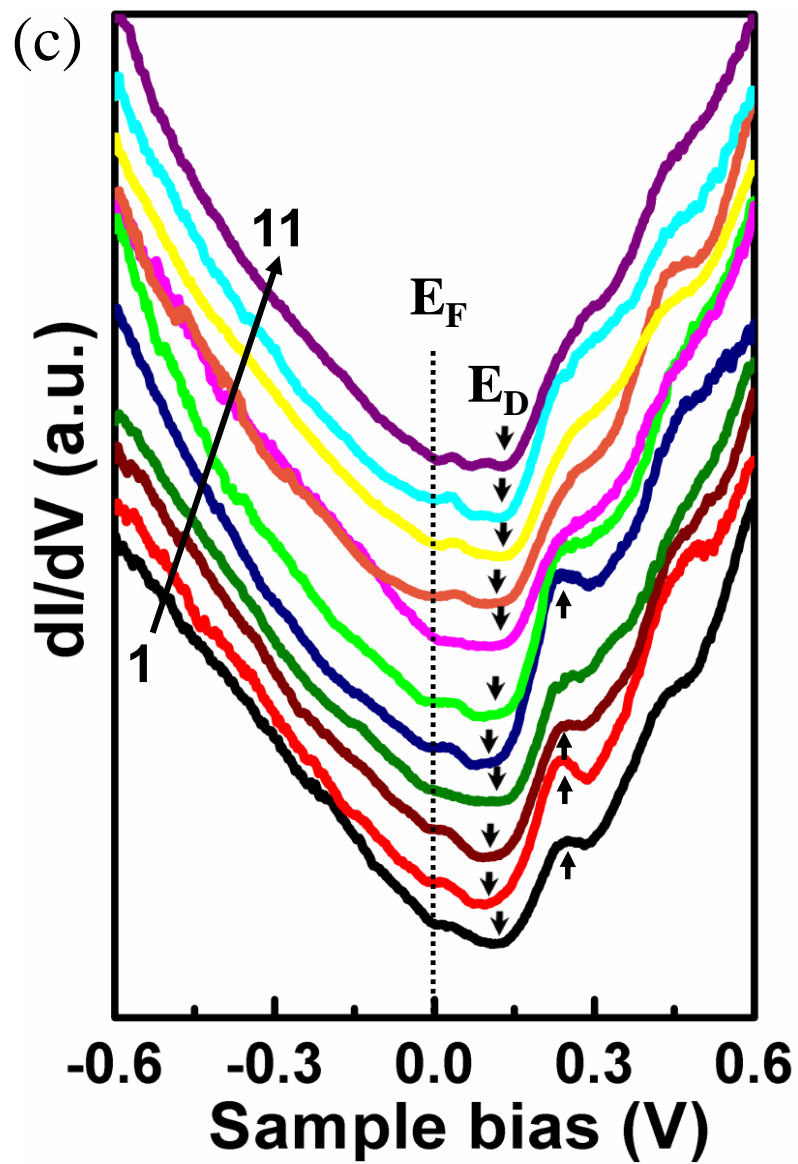

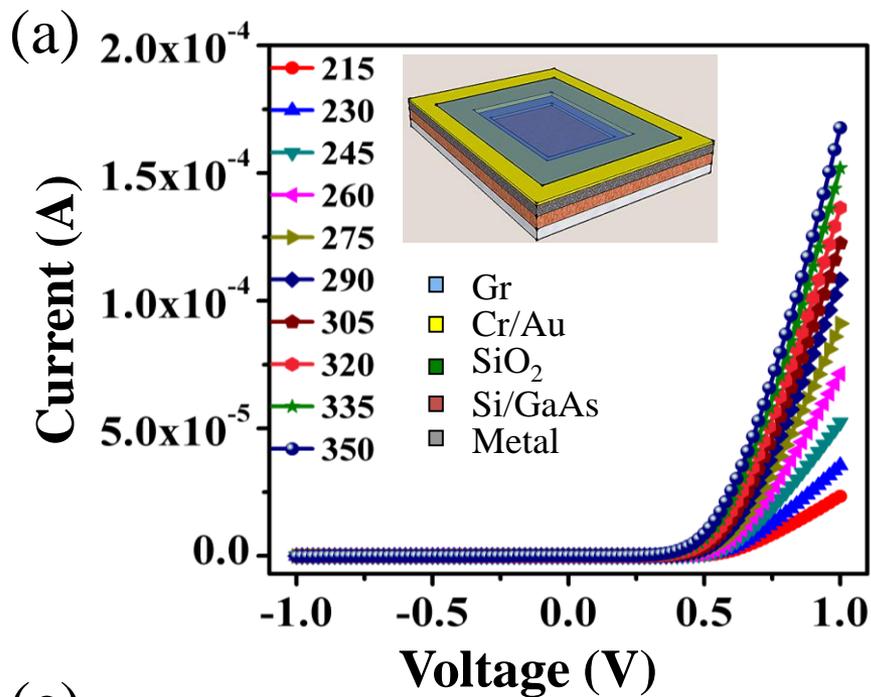 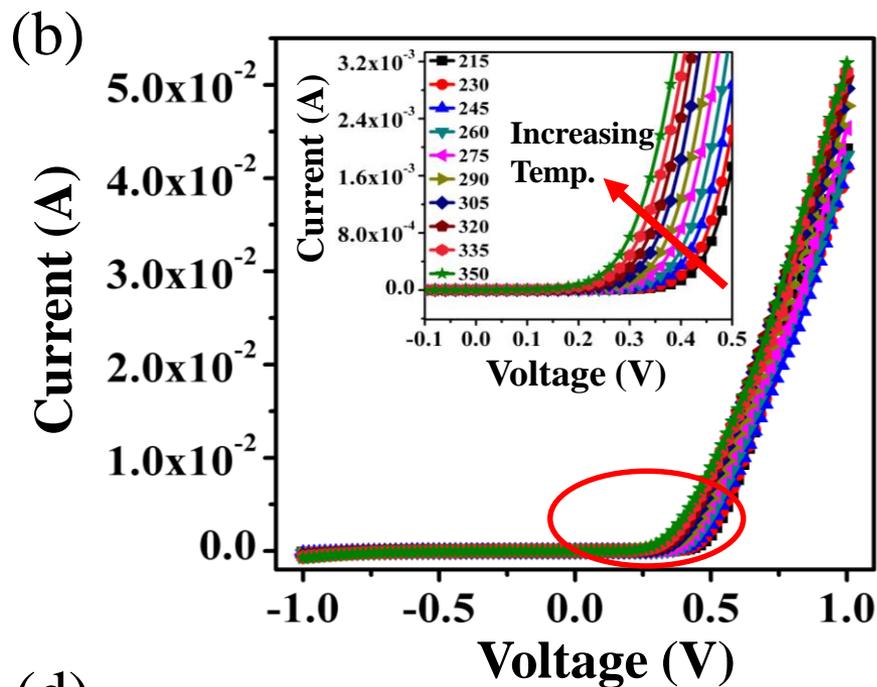
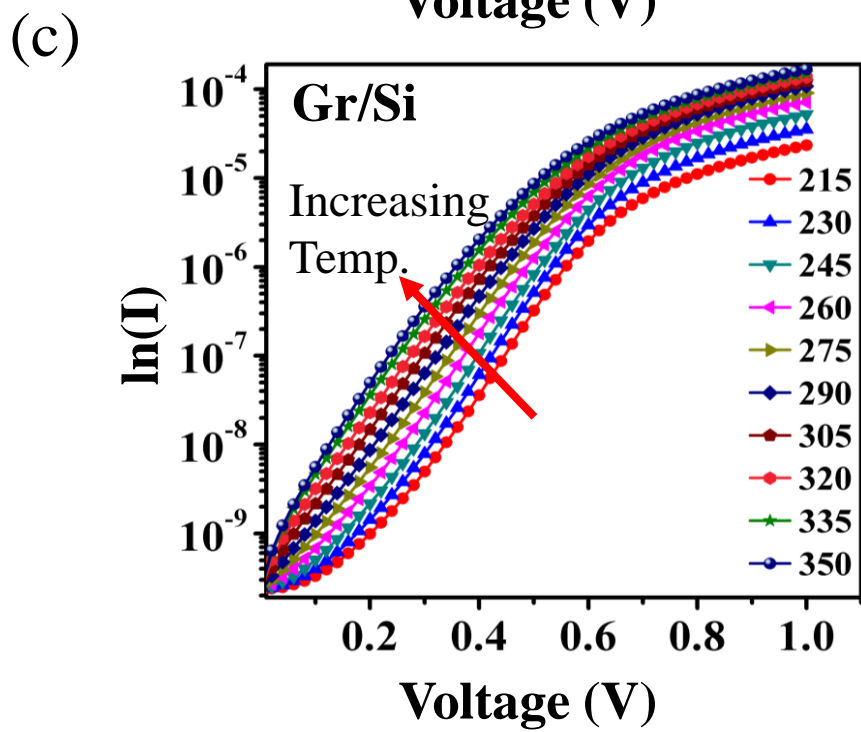 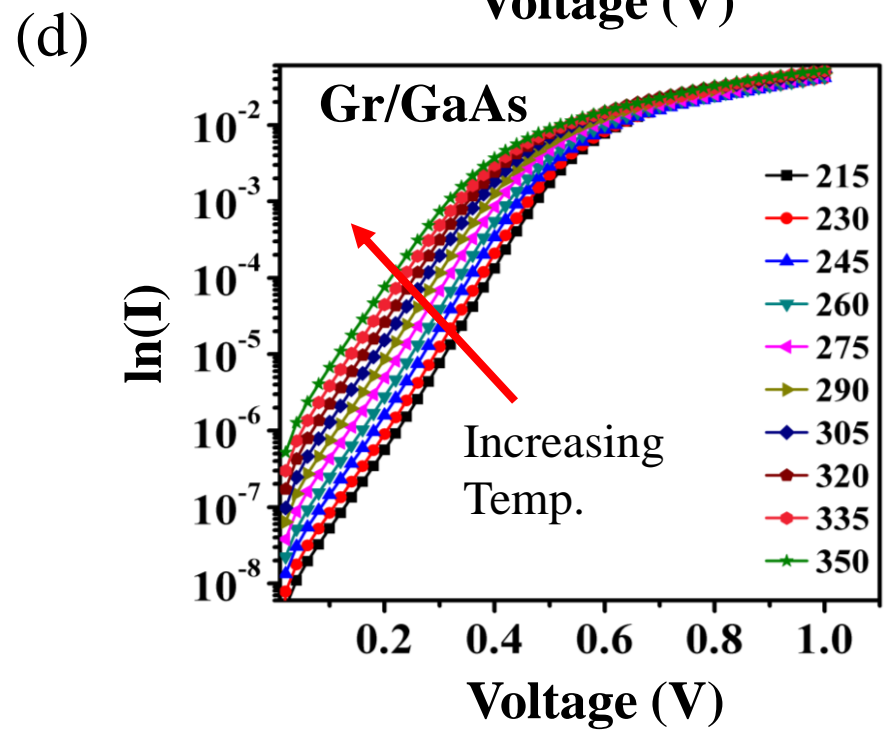

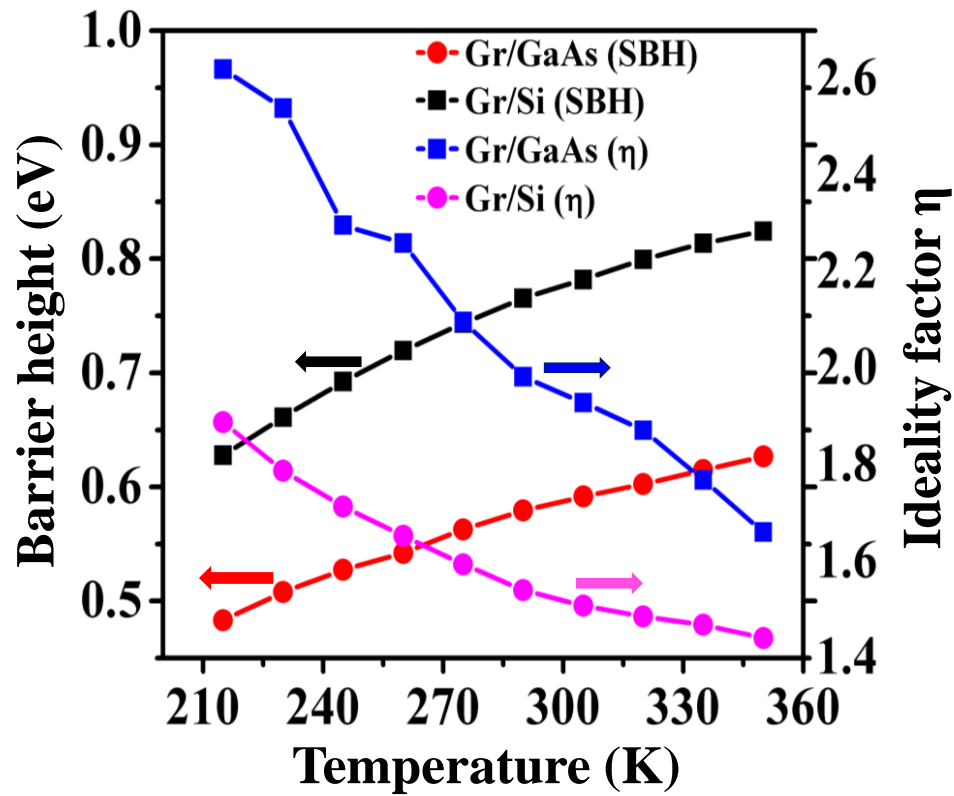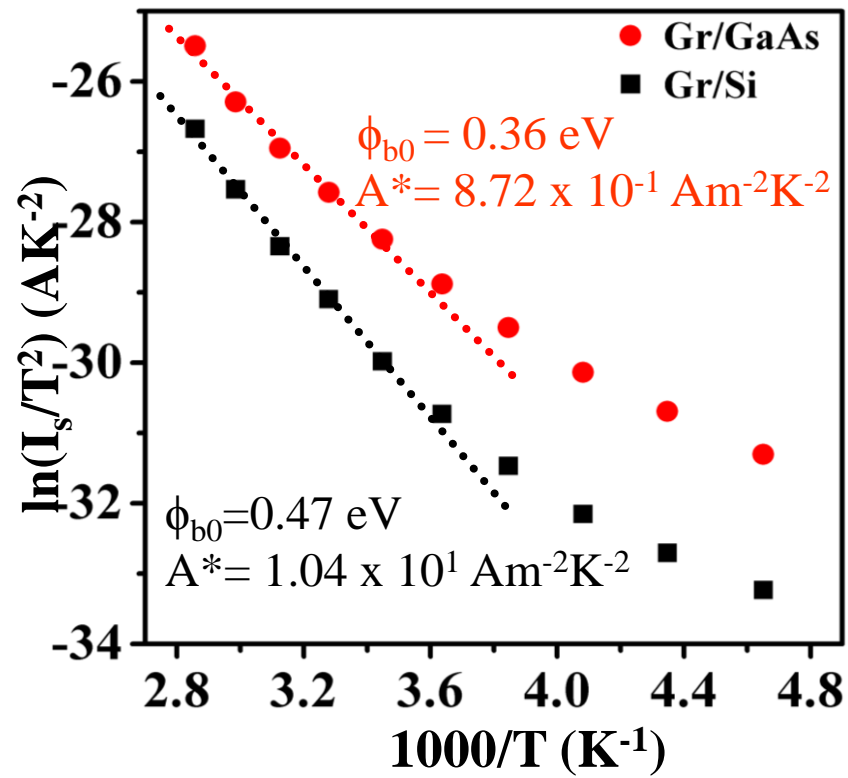

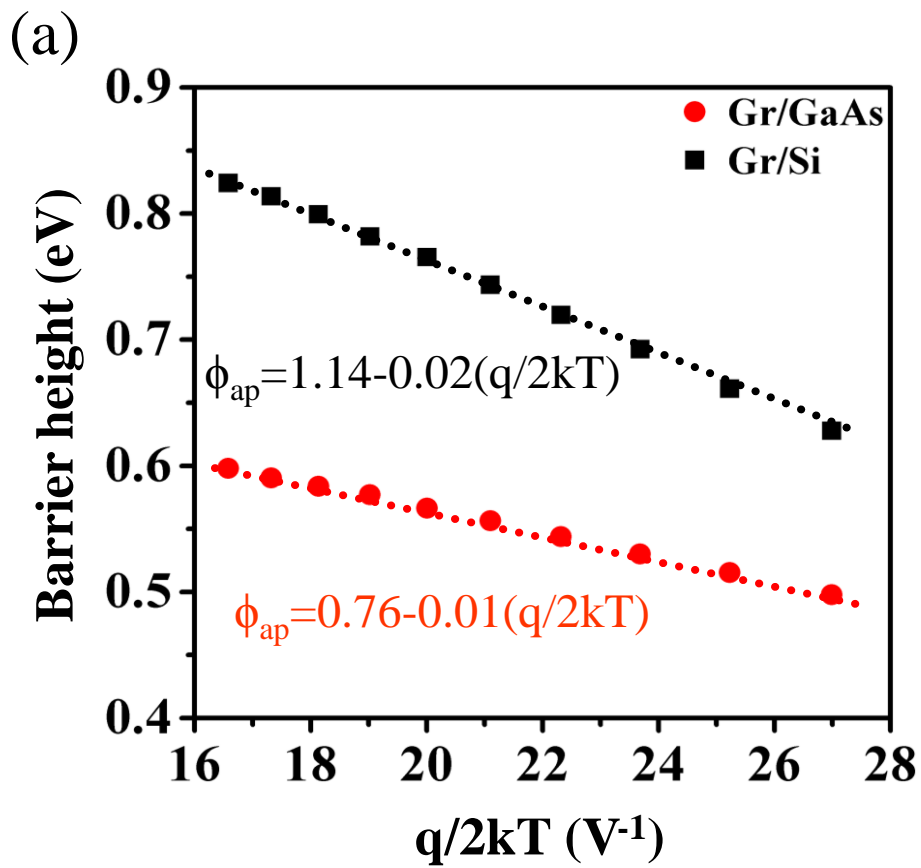 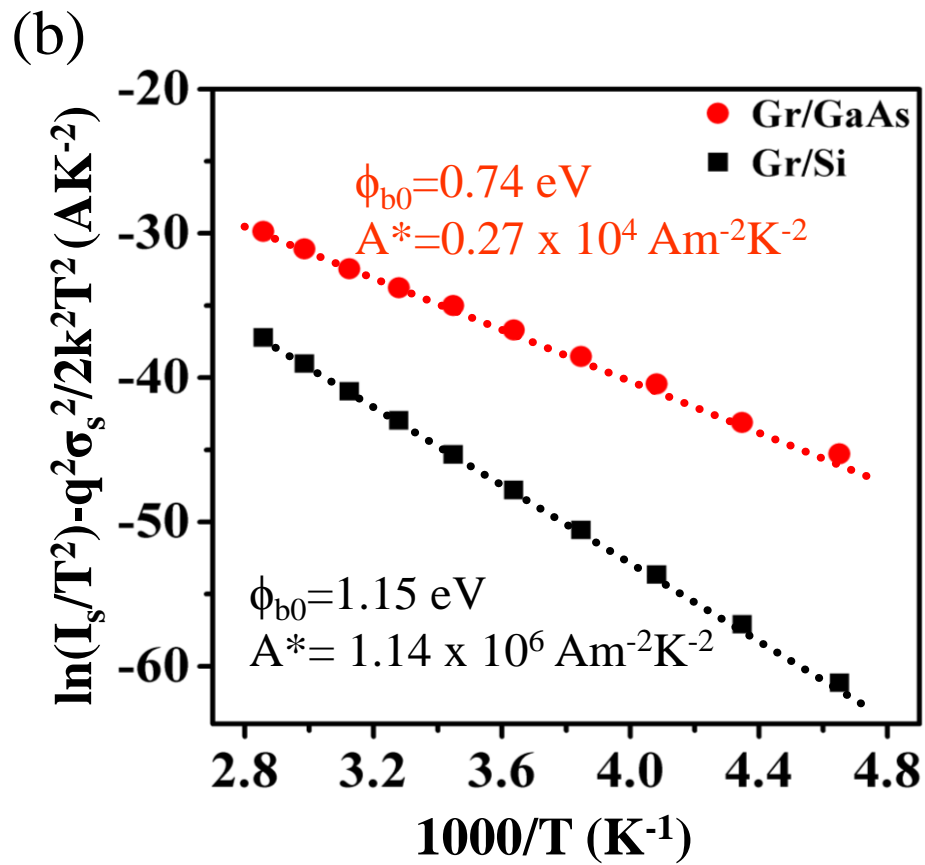

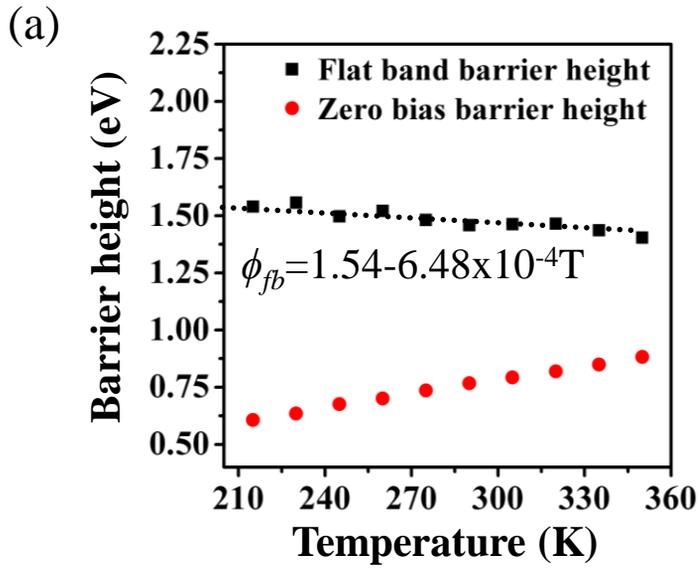

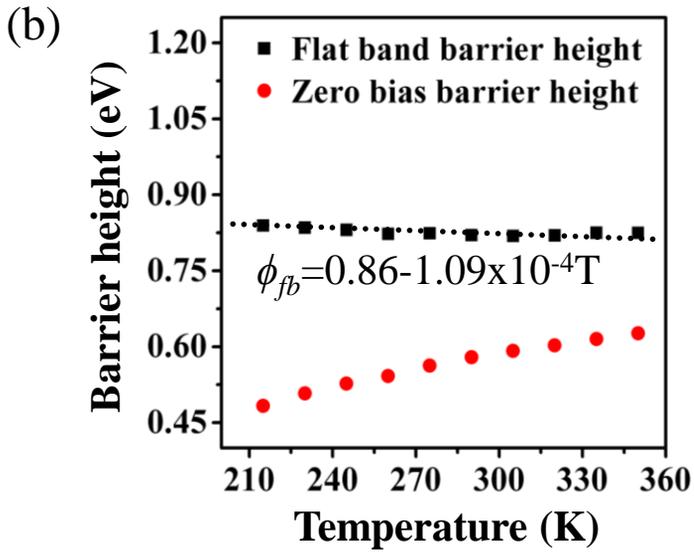

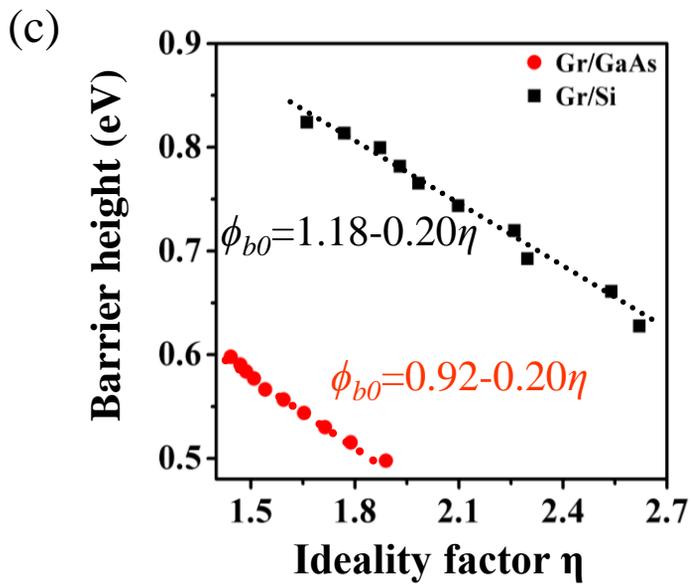